\documentclass[openacc]{rstransa}

\newcommand{\edit}[1]{{#1}}
\newcommand{\CG}[1]{{#1}}

\titlehead{Research}

\begin{document}

\title{Modes of the Dark Ages 21cm field accessible to a lunar radio interferometer}

\author{
Philip Bull,$^{1,2}$ Caroline Guandalin,$^{3}$ and Chris Addis$^{3}$}

\address{$^{1}$Jodrell Bank Centre for Astrophysics, University of Manchester, Manchester M13 9PL, UK\\
$^{2}$Department of Physics and Astronomy, University of Western Cape, Cape Town 7535, South Africa\\
$^{3}$School of Physical and Chemical Sciences, Queen Mary University of London, London E1 4NS, UK
}

\subject{astrophysics}


\corres{Philip Bull\\
\email{phil.bull@manchester.ac.uk}}

\begin{abstract}
At redshifts beyond $z \gtrsim 30$, the 21cm line from neutral hydrogen is expected to be essentially the only viable probe of the 3D matter distribution. The lunar far-side is an extremely appealing site for future radio arrays that target this signal, as it is protected from terrestrial radio frequency interference, and has no ionosphere to attenuate and absorb radio emission at low frequencies (tens of MHz and below). We forecast the sensitivity of low-frequency lunar radio arrays to the bispectrum of the 21cm brightness temperature field, which can in turn be used to probe primordial non-Gaussianity generated by particular early universe models. We account for the loss of particular regions of Fourier space due to instrumental limitations and systematic effects, and predict the sensitivity of different representative array designs to local-type non-Gaussianity in the bispectrum, parametrised by $f_{\rm NL}$. Under the most optimistic assumption of sample variance-limited observations, we find that $\sigma(f_{\rm NL}) \lesssim 0.01$ could be achieved for several broad redshift bins at $z \gtrsim 30$ if foregrounds can be removed effectively. These values degrade to between $\sigma(f_{\rm NL}) \sim 0.03$ and $0.7$ for $z=30$ to $z=170$ respectively when a large foreground wedge region is excluded.
\end{abstract}





\maketitle


\section{Introduction}

The cosmic Dark Ages, roughly corresponding to the redshift range $z \approx 30 - 1000$ \cite{2001PhR...349..125B, 2008PhRvD..78j3511P}, constitute one of the last great frontiers of observational astronomy. This period is sandwiched between the recombination era, when the baryonic content of the Universe became electrically neutral for the first time ($z \simeq 1090$), and the Cosmic Dawn ($z \sim 15 - 30$), when the first stars and galaxies formed and began to reionise the neutral intergalactic medium (IGM).

The large lookback time and lack of luminous sources from this period make any form of direct observation of the Dark Ages challenging. The baryonic matter at this time comes predominantly in the form of neutral Hydrogen (HI) however, which has a distinctive spectral line deep in the radio part of the spectrum, at a wavelength of 21.1 cm (1420.4 MHz) \cite{2012RPPh...75h6901P}. By the time of observation today, this has been strongly redshifted into the low-frequency end of the radio spectrum, at tens of MHz and below. This is quite fortunate, as the late Universe is optically thin to these frequencies. Radio emission from the Dark Ages can travel, effectively unimpeded, from its time of emission until the present day, making it a highly promising probe of this otherwise `dark' epoch \cite{2012RPPh...75h6901P}.

Immediately following recombination at $z \simeq 1090$, fluctuations in the baryonic matter density, including the neutral hydrogen gas, had a different amplitude and scale dependence compared with the dominant cold dark matter (CDM) component \cite{2007ApJ...662....1P}. Over time the baryon distribution evolved to match the CDM distribution on sub-horizon scales, down to around the Jeans scale ($k_J \approx 300$~Mpc$^{-1}$), where baryon pressure effects take over \cite{2007ApJ...662....1P, munoz2015}. By measuring the statistical clustering properties of the baryons, we can therefore infer how the dark matter is clustered, which in turn can be related back to the cosmic initial conditions set during the inflationary epoch.

The 3D spatial distribution of the neutral hydrogen is not observed directly however. The actual observable quantity is the intensity, or equivalently the brightness temperature, $T_b$, of the HI as a function of angle on the sky and radio frequency. The frequency can be mapped to an observed redshift, which allows us to reconstruct a 3D map of the HI as it is projected onto our past lightcone, i.e. the frequency dimension captures the evolution of the brightness temperature field in both time and comoving distance from us. The peculiar motions of the gas also contribute a Doppler shift, further distorting the distribution in the frequency direction \cite{2012RPPh...75h6901P}. These effects are expected to be relatively mild, and predictable. 

Dark matter clustering can also be studied using large-scale structure probes at later times. The tracer populations used there are expected to have a complex relationship with the dark matter distribution that they are embedded in however -- the connection between galaxies and the dark matter field is difficult to model, even in light of modern hydrodynamical simulations, and so contributes substantial theoretical uncertainties into the interpretation of the observed clustering \cite{2018ARA&A..56..435W}. This is particularly the case on smaller scales, of order Mpc and below, where \edit{dark matter has assembled into collapsed halo objects} that are populated by different types of galaxies according to complex non-linear galaxy formation and feedback processes \cite{2023arXiv230308752A}. The effects of non-linear gravitational collapse influence increasingly large scales as time goes on, breaking the connection between the observed clustering of matter and the primordial fluctuations that initially seeded it.

The picture is somewhat simpler in the Dark Ages. Non-linear collapse is confined to much smaller scales -- large collapsed objects have not yet had time to form -- and there are not yet any galaxies to participate in complicated formation and evolution processes. Nor do we need to worry about complicated and uncertain radiative processes that affect the ionisation properties of the gas during Cosmic Dawn and the subsequent Epoch of Reionisation (EoR) between $z \sim 6 - 30$. The 21cm brightness temperature is not simply a linear tracer of the baryon density or CDM density however. The local brightness temperature fluctuation $\delta T_b$ depends on the thermal state of the gas (via the spin temperature, $T_s$) as well as on the local HI density, and is also modulated by an optical depth term that depends on the line-of-sight peculiar velocity \cite{2007ApJ...662....1P, munoz2015}. While these terms and their relation to the underlying baryon and CDM fluctuations, $\delta_b$ and $\delta_c$, can be calculated analytically, the mapping between them necessarily includes terms beyond linear order.

For cosmological interpretation, the quantities of interest are not the 3D matter field itself, but its statistical properties, as these can be predicted theoretically, regardless of the particular statistical realisation of the field that we observe. Existing observational constraints, e.g. from Cosmic Microwave Background (CMB) experiments, point towards a cosmic matter distribution that is approximately statistically homogeneous, isotropic, and close to Gaussian-distributed \cite{2020A&A...641A...7P}. The implication of these properties is that, to a very good approximation, the statistics of the matter distribution are fully described by the power spectrum of the matter density fluctuations, i.e. their 2-point function in Fourier space, defined by
\begin{equation}
\langle \delta_{\rm m}(\vec{k}) \delta_{\rm m}^*(\vec{k})^\prime \rangle = (2\pi)^3 \delta^{(3)}(\vec{k} - \vec{k}^\prime) P_{\rm m}(|\vec{k}|).
\end{equation}
Here, the Fourier-space matter density fluctuation is defined through $\rho_m(\vec{k}) = \bar{\rho}_m \left ( 1 + \delta_{\rm m}(\vec{k})\right )$, where $\bar{\rho}_m$ is the mean matter density at a given redshift.
The angle brackets denote an ensemble average (which can be replaced by a spatial average over sufficiently large volumes), $\delta^{(3)}$ is the 3D Dirac delta function, and $P_{\rm m}(|\vec{k}|)$ is the total matter power spectrum, which encodes the variance of the Fourier-space baryon + CDM field as a function of wavenumber (inverse distance scale), $k \equiv |\vec{k}|$. Under the assumption of Gaussianity, all even higher-order moments of the matter distribution can be calculated from products of the power spectrum, while all odd moments vanish. Exact Gaussianity is broken by non-linear gravitational processes however, which induce higher-order moments and couple different Fourier modes. Importantly, mild non-Gaussianities can also be caused by primordial processes, such as the differential evolution of the fields in multi-field inflation models \cite{2010AdAst2010E..72C}. The detection of primordial non-Gaussianity (PNG) in the cosmic matter distribution would therefore give us a uniquely valuable way of probing at least some of the physical processes at work in the very earliest moments of the Universe's history.

\edit{A useful probe} of non-Gaussianity is the bispectrum, i.e. the 3-point function of the matter density field in Fourier space, which vanishes for a perfectly Gaussian field. Different types of primordial model predict different amplitudes and shapes of the bispectrum \cite{liguori2010}. The most commonly studied is local-type non-Gaussianity, which gives rise to a bispectrum that is largest for `squeezed' 3-point configurations (i.e. where the triangle formed by the three $\vec{k}$ vectors of the 3-point function are elongated isosceles triangles, $k_1 \simeq k_2 \gg k_3$). The amplitude of this bispectrum shape is typically parametrised by $f_{\rm NL}$, and current constraints from the CMB place an upper limit of approximately $|f_{\rm NL}| \lesssim 5$ \cite{2020A&A...641A...9P}. Depending on the inflation model, one can obtain values of $f_{\rm NL} \sim 1$ (for particular multi-field models for instance), down to a strong prediction of $f_{\rm NL} \sim 10^{-3}$ for single field models \cite{2003JHEP...05..013M, 2014arXiv1412.4671A}. Other bispectrum shapes can also be obtained and their amplitudes constrained \cite{2013JCAP...03..037C, 2013JCAP...09..026B}.

The primary CMB temperature fluctuations have been measured well enough that no further (substantial) improvement in bispectrum constraints can be made from this source -- the constraints are now dominated by `cosmic variance', which is the intrinsic uncertainty due to having only a finite number of samples of a given quantity (in this case, a finite number of observable Fourier modes in the observed CMB). Late-time large-scale structure probes measure a larger number of Fourier modes than the CMB, and so can be used to further improve constraints and push closer to the \edit{$\sigma(f_{\rm NL}) \lesssim 1$ level} required to test some types of inflationary model, albeit with the difficulty of needing to account for non-linearity and \edit{astrophysical modelling uncertainty \cite{2014arXiv1412.4671A, 2014arXiv1412.4872D, 2016JCAP...06..014T, 2017PhRvD..95l3513D, 2023EPJC...83..320J}.}

Pushing to the $f_{\rm NL} \sim 10^{-3}$ level requires vastly more Fourier modes however \cite{2021RSPTA.37990561S}. This is where Dark Ages 21cm mapping experiments have a crucial role to play. There are very many more modes available from extending to higher wavenumbers than are reasonably accessible to low-redshift surveys, as the number of Fourier modes in a survey volume scales like $k_{\rm max}^3$, where $k_{\rm max}$ is the maximum recoverable Fourier wavenumber. Furthermore, the non-primordial contributions to the Dark Ages 21cm bispectrum, such as those due to the non-linear mapping between the 21cm signal and the underlying CDM field, can be calculated analytically, with comparatively fewer theoretical uncertainties in how to model them.

In this article, we examine some of the practical challenges associated with observing the bispectrum of 21cm brightness temperature fluctuations from the cosmic Dark Ages. The highest redshifts (and a large fraction of the Fourier modes) can only be accessed by radio telescopes outside the Earth's atmosphere, due to the scattering effect of the ionosphere on low-frequency radio waves, while radio frequency interference from around the Earth is also challenging to identify and remove at the relevant frequencies. Hence, there have been a number of proposals to build a Dark Ages 21cm instrument on the far side of the Moon, or in lunar orbit.

We use the technique of Fisher forecasting to obtain simplified predictions for how well different array configurations should be able to measure the HI bispectrum. We incorporate the effect of losing modes to systematic effects such as radio foreground emission in our forecasts, \edit{as well as the intrinsic limitations due to instrumental resolution.}

\begin{figure}[!htb]
\includegraphics[width=0.85\linewidth]{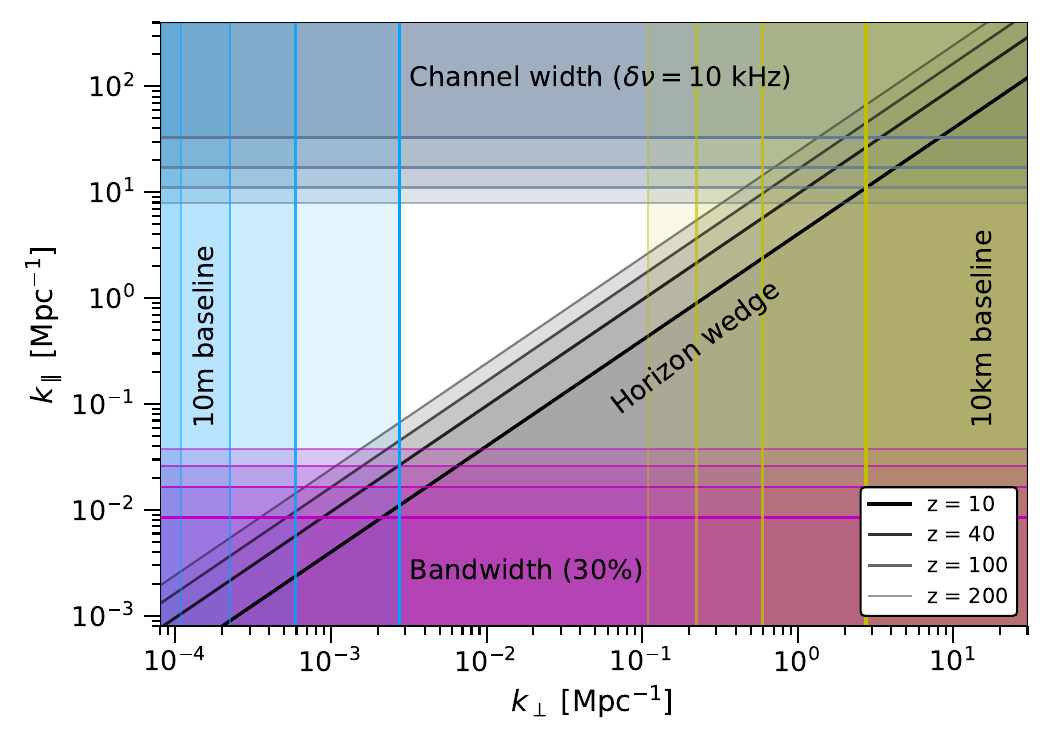}
\caption{`Exclusion' plot showing which cylindrical Fourier modes ($k_\perp, k_\parallel$) are observable with an interferometer for four representative redshifts. Shaded regions would be excluded from the observations due to various observational effects: the foreground wedge (grey, diagonal); the frequency channel width (grey-blue, top); the fractional bandwidth of the observation (magenta, bottom); and the minimum and maximum baseline lengths (blue, left and yellow, right respectively). We have assumed representative numbers here: 10~kHz frequency channels, a 30\% effective bandwidth at each redshift, and minimum and maximum baselines of 10m and 10km respectively.}
\label{fig:scales}
\end{figure}

\section{Lunar radio array configurations} \label{sec:config}

Leaving matters such as deployment, power, data transmission, and array calibration aside, the main properties that determine the observing characteristics of a radio interferometer array are its frequency range (bandwidth) and spectral resolution, the field of view of individual receiving elements (also called the primary beam), and the distribution of baselines, i.e. the number of available correlated antenna pairs as a function of the length and orientation of the separation vector between them. All of these terms are represented in the noise power spectrum, $P_N$ (see Eq.~\ref{eq:PN} below). In this section, we attempt to distill the various lunar arrays that have been proposed into a handful of model configurations that we can then use to produce representative forecasts.

\subsection{Fourier-space sampling function} \label{sec:sampling}

First, we briefly review the connection between the instrumental configuration, including the baseline distribution, and the set of Fourier modes on the sky that can be observed by the interferometer and thus included in the bispectrum measurements. An illustration for some representative instrumental properties is shown in Fig.~\ref{fig:scales}.

\paragraph{Baseline distribution} The baseline distribution can be represented as a number density of baselines in the $uv$-plane. Vectors in this plane can be mapped into transverse Fourier vectors by $\vec{k}_\perp = 2\pi \vec{u} / \chi(z)$ (where $\chi$ is the comoving distance to redshift $z$), assuming that $\vec{k}_\perp$ has been defined in a plane on the sky that is parallel to the plane of the interferometer array, which we take to be perfectly flat, tangent to the lunar surface, with a normal that points at the zenith. Under these assumptions, the $uv$-plane baseline distribution can be calculated by finding the vector $\vec{u} = \vec{d} / \lambda$ corresponding to each baseline, where $\vec{d}$ is the separation vector between the antennas. The set of baseline vectors can then be binned appropriately in the $uv$-plane to obtain the number density $n_b$. Note that the $uv$-plane distribution changes with observing wavelength since $\vec{u} \propto 1/\lambda$, and so $n_b(\vec{u})$ will also vary with redshift.

In the flat-sky limit, each baseline can be thought of as being sensitive to the amplitude (in brightness temperature) of a single Fourier mode on the sky with wavevector $\vec{k}_\perp$. The field of view of the antennas acts as a window function on the set of Fourier modes that the array is sensitive to however, and so introduces a correlation scale in the $uv$-plane of approximately $\Delta u \sim 1 / \sqrt{\Omega_{\rm FOV}}$. This scale can be used as the minimum bin width for the baseline number density distribution, and also sets a lower limit on the minimum recoverable $k_\perp$ mode (i.e. maximum recoverable angular scale).

Importantly, only Fourier modes on the sky represented by baselines that are present in the array can be recovered. Unrepresented Fourier modes are not measured, and therefore cannot be used in the bispectrum measurements etc. In Earth-based interferometry applications, it is common to use the rotation of the baseline vectors with respect to the sky, due to the Earth's diurnal rotation, to perform {\it rotation synthesis}. As the Earth rotates, each baseline migrates along an elliptical track in the $uv$-plane as its orientation and projection of the baseline changes with respect to a reference point on the sky. This allows a wider range of Fourier modes to be recovered by each baseline as the rotation progresses.

For lunar applications, the situation is different, as there is no diurnal rotation. The Moon orbits the Earth during the lunar month, and the Earth orbits the Sun, so observations taken across a terrestrial year will allow some degree of rotation synthesis to be achieved \cite{1991ASPC...19..420B}. Different patches of the sky will also rise and set throughout the month. Since the Sun is a bright radio source, observations would normally be taken during the lunar night however, and so this prevents some segments of the tracks in the $uv$-plane from being recovered. 

\paragraph{Foreground wedge} Radio interferometer visibilities measure the integrated sky intensity distribution after it has been modulated by a baseline-dependent fringe pattern. To a good approximation, the fringe pattern for each baseline can be mapped to a transverse Fourier mode, $k_\perp$, at each frequency. The sky intensity distribution is also modulated by the primary beam pattern of the antennas however, which also depend on frequency, but in a different way. When a Fourier transform is performed in the frequency direction, i.e. to give the visibility in terms of radial Fourier mode, $k_\parallel$, this additional modulation gives rise to a coupling between transverse and radial modes of the sky signal. This scatters intrinsically spectrally-smooth sky emission (i.e. the foregrounds) at low-$k_\parallel$ into an extended wedge-shaped region in ($k_\perp, k_\parallel$) space. The maximum extent of the wedge region is in principle related to the maximum geometric delay in arrival time of a wavefront between the two antennas of a baseline, which occurs when a source is on the horizon. This `worst case' foreground-contaminated region is referred to as the `horizon wedge'.

The intensity of the contamination can vary within the wedge, depending on the shape of the primary beam pattern -- for dipole antennas the wedge is quite evenly contaminated, for instance, while for parabolic reflectors, it tends to be more localised into the `prongs' of a pitchfork shape, with cleaner regions in between them. Since most lunar array concepts employ dipoles for reasons of cost and simplicity, we can anticipate severe contamination throughout the entire wedge region. Cleaning foreground emission from inside the wedge region requires exceptionally good models of both the primary beams and the foreground emission however, and many ground-based arrays choose a more conservative `avoidance' strategy instead. This is where 21cm signal modes within the wedge region are assumed to be irretrievable and so are excised, while a variety of analysis choices are made to prevent power from leaking out of the wedge region into an otherwise clean `window' region.

In the absence of a detailed plan to support extraction of 21cm signal modes within the wedge region by characterising the foregrounds and beam patterns of a lunar array with extreme precision, \edit{we can conservatively assume that modes within the horizon wedge are unusable for 21cm cosmology (see Fig.~\ref{fig:scales}). More optimistically, calibration methods such as \cite{nucal} may allow substantial suppression of the foregrounds however. We bracket the possibilities by producing forecasts for both horizon-wedge (pessimistic) and no-wedge (optimistic) scenarios.}

\paragraph{Bandwidth and spectral resolution} The frequency axis of the observations simultaneously encodes the evolution of the signal with redshift, $z$, and the variations of the 21cm field in the radial or line-of-sight direction (represented by the radial Fourier wavenumber, $k_\parallel$). It is common to make an approximation that the redshift evolution is negligible over a small frequency range (i.e. within a sufficiently narrow redshift bin), so that frequency channels can be mapped to a radial distance within a 3D volume at a `fixed' central redshift. The frequency resolution then gives the maximum radial Fourier wavenumber $k_\parallel^{\rm max}$ that can be measured in the volume. Similarly, the bandwidth over which the redshift evolution is neglected can be converted into a maximum radial extent of the 3D volume, and so sets the fundamental radial mode, $k_\parallel^{\rm min}$. 

While there are methods that allow this redshift-binning approximation to be avoided, working with the Fourier-space power spectrum and bispectrum typically requires some kind of redshift binning to be done, and so we take it as an unavoidable aspect of our analysis here. The redshift bin width can be chosen to keep the cosmological evolution of the signal across the bin sub-dominant, and so will typically vary with frequency. As a somewhat maximal choice, we assume a 30\% fractional bandwidth within each bin. For bins centred at $z=10, 40, 100, 200$ ($\nu \simeq 130, 35, 14, 7$~MHz), this would equate to bandwidths of $\Delta\nu \simeq 40, 10, 4, 2$~MHz.

\begin{table}[t]
  \begin{tabular}{|p{2cm}|c|c|c|c|}
  \hline
  {\bf Configuration} & {\bf No. antennas} & {\bf Approx. area} & {\bf Freq. range} & {\bf Redshift range} \\
  \hline
  Stage I & $\mathcal{O}(500)$ & 1~km $\times$ 1~km & 20 -- 60~~MHz & 23 -- 70~\, \\
  Stage II & $\mathcal{O}(10^4)$ & 5~km $\times$ 5~km & 10 -- 60~MHz & 23 -- 140 \\
  Stage III & $\mathcal{O}(10^5)$ & 10~km $\times$ 10~km & ~5 -- 60~MHz & 23 -- 280 \\
  \hline
  \end{tabular}
  \caption{Representative specifications of the three development stages for lunar 21cm arrays. These numbers are loosely aligned with mission concepts in the literature, including ones that propose a staged approach.} \label{tab:expts}
\end{table}

\subsection{Representative baseline distributions} \label{sec:baselines}

\begin{figure}[!th]
\centering
\includegraphics[width=0.8\linewidth]{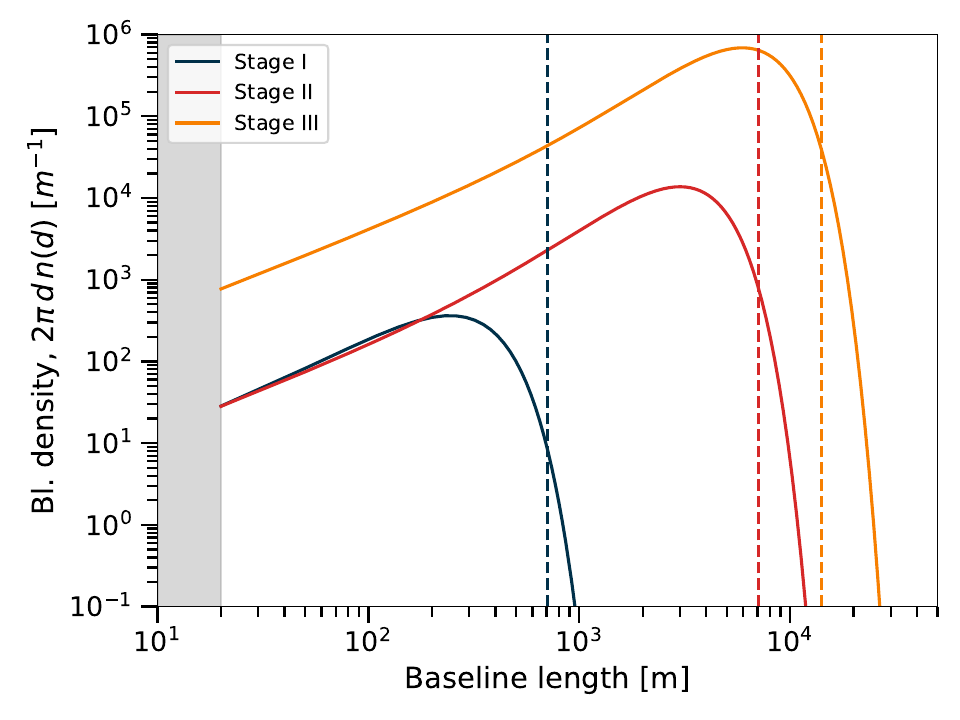}
\caption{Notional baseline distributions for the three representative arrays. The functional form is chosen to have a similar form to the one for FarView for the Stage~III experiment. Vertical dashed lines show the maximum baseline length for a regular square array with linear extent 500m, 5km, and 10km respectively. The Stage I and II experiments have similar densities for the shortest baselines, differing only in the number of long baselines, consistent with extending a dense core by adding outriggers at intermediate and large distances.}
\label{fig:bldist}
\end{figure}

In this section, we attempt to distill the various mission concepts in the literature into a set of representative development stages for a lunar 21cm array. We have used the CoDEX \cite{2019arXiv190804296K}, FarSide/FarView \cite{2021arXiv210308623B, farview}, and ROLSS/DALI \cite{2007astro.ph..1770L} concepts as the basis for the following.

\paragraph{Stage I} A relatively small pathfinder array with a dense core, likely in a grid layout that is the minimum needed to test antenna deployment technologies and implement an FFT-based correlator, e.g. a $16 \times 16$ ($=256$ antennas) or $32 \times 16$ ($=512$ antennas) array. To soften the technical requirements, choices such as a reduced bandwidth (e.g. $20 - 60$ MHz) and relatively short maximum baseline length can be made. For maximum and minimum baseline lengths of $\sim1$~km and $30$m respectively, scales in the range $5\times 10^{-3} \lesssim k_\perp \lesssim 5\times 10^{-2}~{\rm Mpc}^{-1}$ would be accessible, i.e. around the top end of the BAO scale. 

\paragraph{Stage II} A large array of around 10,000 antennas with a large core plus some outriggers to increase the maximum baseline length to around 5~km. Building on the technology demonstrated by the Stage I instrument, this could allow for deviations from an FFT-friendly regular grid into a more balanced layout, permitting longer baselines for imaging. Improvements in data rate, antenna design etc. would allow a wider bandwidth to be observed, perhaps in the $10 - 100$~MHz range, covering both Cosmic Dawn and well into the Dark Ages, $13 \lesssim z \lesssim 140$ (note that we only consider $z \gtrsim 20$ in this paper). Angular scales in the range $5\times 10^{-4} \lesssim k_\perp \lesssim 5\times 10^{-1}~{\rm Mpc}^{-1}$ would be accessible with this instrument, permitting some initial imaging applications at the lowest redshifts.

\paragraph{Stage III} A giant array of 100,000 or more antennas over a large area of $10~{\rm km} \times 10~{\rm km}$ or so. This would further improve on the frequency range of Stage II, probing frequencies down to 5~MHz or so, and angular scales in the range $10^{-4} \lesssim k_\perp \lesssim 10^{0}~{\rm Mpc}^{-1}$.

In all cases, we employ a minimum baseline length of 20m. This should be compared with the maximum wavelength of between $\sim 15 - 60$m (at 20~MHz and 5~MHz respectively), and a typical dipole length of order $\sim 10$m. Neighbouring antennas will be within the near-field of one another at higher frequencies, and we can expect mutual coupling effects to be important.

Rather than defining an explicit baseline distribution for each Stage, we attempt to capture representative distributions based on a fitting function for the circularly-averaged density,
\begin{equation}
n(d) = \mathcal{A} \left ( D - D_0 \right )^2 \exp \left [ -\left ( \frac{D - D_0}{w} \right )^2 \right ],
\end{equation}
\edit{where $\mathcal{A}$ is a normalising constant determined by the total number of unique baselines, $\int_{D_{\rm min}}^{D_{\rm max}} 2\pi D\, n(D) dD = N_{\rm ant}(N_{\rm ant}-1) / 2$}, where the number of antennae is given in Table~\ref{tab:expts}. The function was designed to give a good fit to the FarView concept's baseline distribution, which has a regular grid in the core and semi-random outlier stations \cite{farview}. For Stages I, II, and III we chose the parameters $D_0 = -180, -1000, -2100$~m and $w = 310, 3100, 6200$~m respectively. These values give maximum baseline lengths around the right value for the specifications in Table~\ref{tab:expts}. The resulting baseline distributions are shown in Fig.~\ref{fig:bldist}.

\section{Fisher forecasting formalism} \label{sec:fisher}

In this section, we describe the theoretical calculation of the 21cm power spectrum and bispectrum during the dark ages, and the Fisher forecasting formalism that we use. 
We base our calculations on \cite{munoz2015} and \cite{2020JCAP...11..052K}.

The Fisher matrix for the anisotropic bispectrum can be calculated as
\begin{equation}\label{eq:fisher}
    F_{\alpha \beta}^{(B)}(z_i) = \frac{1}{4\pi} \sum_{\rm triangles} \int_{-1}^{+1} {\rm d}\mu_1 \int_0^{2\pi} {\rm d}\phi\, \frac{1}{\Delta B^2_{\rm HI}(z_i)} \frac{\partial B_{\rm HI}(z_i)}{\partial p_\alpha} \frac{\partial B_{\rm HI}(z_i)}{\partial p_\beta}, 
\end{equation}
where $B_{\rm HI}(z_i) \equiv B_{\rm HI}(\vec{k_1}, \vec{k}_2, \vec{k}_3, z_i)$ is the 
of the HI brightness temperature in redshift bin $i$ with central redshift $z_i$, and the cosine $\mu_1$ and angle $\phi$ parametrise different orientations of the triangle with respect to the line of sight. \edit{Under the assumption of Gaussianity of the covariance,} the variance of the bispectrum for a given triangle configuration is given by
\begin{equation}\label{eq:deltaB}
    \Delta B_{\rm HI}^2(k_1, k_2, k_3, z_i) = \frac{s_{123}\,\pi}{k_1 k_2 k_3}  \left ( \frac{k_f}{\Delta k} \right )^3 P_{\rm HI}(k_1, \mu_1, z_i)\, P_{\rm HI}(k_2, \mu_2, z_i)\, P_{\rm HI}(k_3,\mu_3, z_i), 
\end{equation}
where $s_{123}$ is a multiplicity factor (equal to 6 for equilateral triangle configurations, 2 for isosceles, and 1 for scalene), and $k_f \equiv 2\pi / L$ is the fundamental mode in Fourier space for an (assumed cubic) survey volume of comoving side length $L$. We will take the continuum limit of Eq.~\ref{eq:fisher} and extend the integrals over the full range of $k_1,k_2$, and $k_3$. For that, we divide Eq.~\ref{eq:fisher} by the total number of equivalent triangles $r_{123}$. Since $r_{123} = (1, 3, 6)$ for equilateral, isosceles and scalene configurations, respectively, $r_{123}\,s_{123}=6$. Hence, we consider the following expression for the Fisher matrix:
\begin{align}
  F^B_{\alpha\beta}&=\frac{V}{4\pi^2}\frac{1}{6}\int d\mu_1 \int d\phi\,\int_{k_{\rm min}}^{k_{\rm max}}\mathrm{d}k_1\,\mathrm{d}k_2\,\mathrm{d}k_3\,\frac{k_1\,k_2\,k_3}{(2\pi)^3}\frac{\partial_{\alpha} B_{\rm HI}(z_i)\,\partial_\beta B_{\rm HI}(z_i)}{P_{\rm HI}(k_1, \mu_1, z_i)\, P_{\rm HI}(k_2, \mu_2, z_i)\, P_{\rm HI}(k_3,\mu_3, z_i)}. \label{eq:fisherbs}
\end{align}
We stress that the continuum approximation may overestimate the total number of triangles used to forecast the errors on $f_{\rm NL}$; in reality, one obtains fewer triangles per configuration due to the binning strategy in Fourier space. For example, one can access fewer triangles with thicker bins, which reduces the expected bispectrum signal \cite{oddo2020}. Therefore, the results presented in Sec.~\ref{sec:results} should be taken as a lower bound for the errors expected with the experiments considered here. We also stress that we are ignoring non-linear corrections to the variance, $\Delta B^2_{\rm HI}$. 

\subsection{Model of the HI bispectrum during the Dark Ages}

Our model for the HI bispectrum is given by (following \cite{munoz2015})
\begin{align}\label{eq:hibispectrum}
    B_{\rm HI}(\vec{k}_1, \vec{k}_2, \vec{k}_3) = B_{\rm HI}^{\rm prim}(\vec{k}_1, \vec{k}_2, \vec{k}_3) + B_{\rm HI}^{\rm grav}(\vec{k}_1, \vec{k}_2, \vec{k}_3) + B_{\rm HI}^{\rm nl}(\vec{k}_1, \vec{k}_2, \vec{k}_3),
\end{align}
where $B_{\rm HI}^{\rm prim}$ is related to the primordial bispectrum of scalar perturbations, $B_\Phi$, as 
\begin{equation}
    B_{\rm HI}^{\rm prim}(\vec{k}_1, \vec{k}_2, \vec{k}_3) = {\cal F}(k_1,k_2,k_3,\mu_1,\mu_2,\mu_3) B_\Phi(k_1,k_2,k_3),
\end{equation}$B_{\rm HI}^{\rm grav}$ is the gravitational contribution to the bispectrum, 
\begin{equation}
    B_{\rm HI}^{\rm grav}(\vec{k}_1, \vec{k}_2, \vec{k}_3) = {\cal G}(\vec{k}_1,\vec{k}_2,\mu_1,\mu_2) P^{\rm L}_{\rm m}(k_1)P^{\rm L}_{\rm m}(k_2) + \text{ 2 perms. },
\end{equation}and $B_{\rm HI}^{\rm nl}$ accounts for the non-linear relation between the brightness temperature and the baryon density:
\begin{equation}
    B_{\rm HI}^{\rm nl}(\vec{k}_1, \vec{k}_2, \vec{k}_3) = {\cal H}(\vec{k}_1,\vec{k}_2,\mu_1,\mu_2) P^{\rm L}_{\rm m}(k_1)P^{\rm L}_{\rm m}(k_2) + \text{ 2 perms. }
\end{equation} 

\begin{figure}[!htb]
\centering
\includegraphics[width=1.05\linewidth]{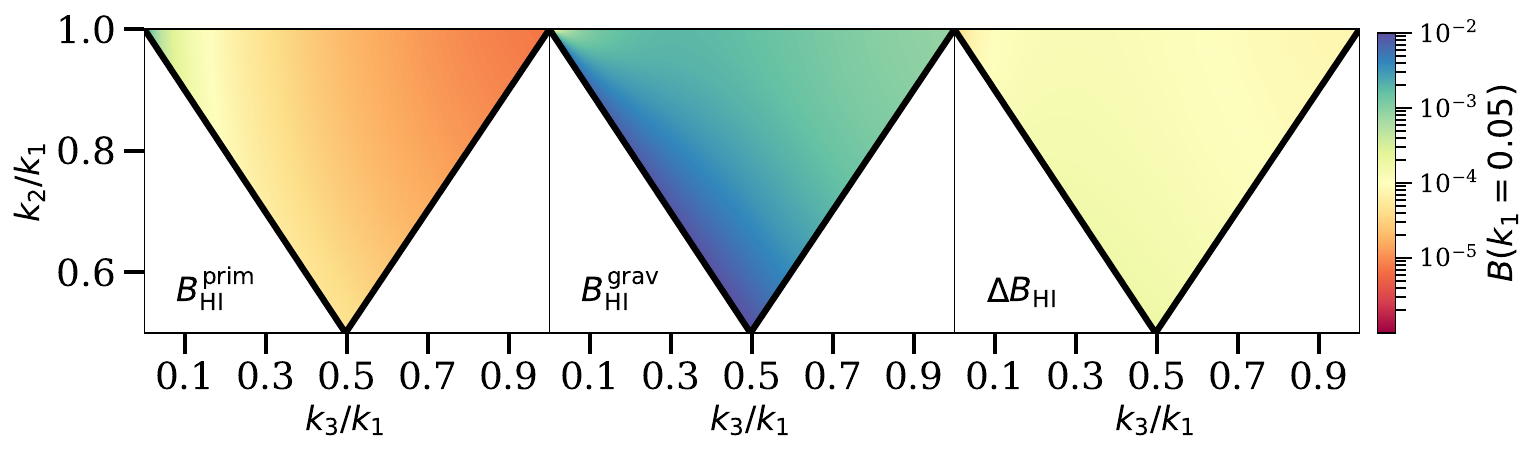}
\caption{\edit{Amplitude of the bispectrum at $z=30$, for triangle configurations where one of the sides is fixed to $k_1 = 0.05$~Mpc$^{-1}$. The left panel shows the primordial bispectrum as a function of triangle configuration, assuming local-type non-Gaussianity with $f_{\rm NL} = 1$ (note the higher amplitude in the top left). The middle panel shows the gravitational contribution to the bispectrum, which is significantly larger. The right panel shows the uncertainty on the bispectrum assuming no thermal noise or instrumental effects (cosmic variance contribution only).}}
\label{fig:triangles}
\end{figure}

The full expressions for these terms are given in \cite{munoz2015}, but we briefly describe them here. The functions ${\cal F}, {\cal G}$ and ${\cal H}$ depend on the angle between the wavevectors $k_i$ and the line of sight $\hat{n}$, $\mu_i = \vec{k}_i \cdot \hat{n}/k_i$, on the mean 21cm brightness temperature at redshift $z$, $\overline{T}_{\rm HI}$, and on the derivative of the brightness temperature $T_{\rm HI}$ with respect to the linear baryon overdensity $\delta_b^{(1)}$: $\alpha \equiv \partial T_{\rm HI}/\partial \delta_b^{(1)}$. For the primordial bispectrum, ${\cal F}$ also carries a dependence on $M(k,z) \equiv 2k^2T(k)D(z)/3\Omega_{\rm m} H_0^2$ (e.g., see \cite{liguori2010}); for the gravitational contribution, ${\cal G}$ also depends on the second-order perturbation theory kernels, $F_2$ and $G_2$, and on $\gamma \equiv \partial T_{\rm HI}/\partial \delta_b^{(2)}$;  and for the non-linear part, ${\cal H}$ also depends on $\beta \equiv \frac{1}{2} \partial^2 T_{\rm HI}/\partial \delta_b^2$.
Finally, the primordial bispectrum $B_\Phi$ can be constructed from a sum of contributions from different bispectrum shapes; we retain only the local-type bispectrum in this work, which can be calculated as
\begin{equation}
B_\Phi^{\rm local}(k_1, k_2, k_3) = 2\, f_{\rm NL} \left[ P_\Phi(k_1) P_\Phi(k_2) + P_\Phi(k_1) P_\Phi(k_3) + P_\Phi(k_2) P_\Phi(k_3) \right].
\end{equation}
Here, $P_\Phi$ is the primordial scalar power spectrum, and $f_{\rm NL}$ is the overall amplitude of the local-type bispectrum, as described above. \edit{Note that \cite{munoz2015} uses the flat-sky approximation and neglects the Wouthuysen-Field effect. An example of some of the different contributions to the bispectrum at $z=30$, including the uncertainty for a cosmic variance-limited scenario, is shown in Fig.~\ref{fig:triangles}.}

In a more comprehensive treatment, we could calculate the Fisher matrix for a range of cosmological parameters 
in order to account for their uncertainties, and correlations between the different parameters. This is largely a matter of evaluating the derivatives of Eq.~\ref{eq:hibispectrum} with respect to these parameters, as they appear in Eq.~\ref{eq:fisher}. In the present article, however, we focus only on the simplest case of an idealised forecast for a single parameter, $f_{\rm NL}$. Since this parameter appears as a prefactor of only a single term in Eq.~\ref{eq:hibispectrum}, the expression for the derivatives simplifies to
\begin{equation}
    \frac{\partial B_{\rm HI}}{\partial f_{\rm NL}} = \frac{\partial B_{\rm HI}^{\rm prim}}{\partial f_{\rm NL}} = {\cal F} \,\frac{\partial B_\Phi^{\rm local}}{\partial f_{\rm NL}}.
\end{equation}
\edit{Implicit in this simplified treatment is an assumption that we are able to perfectly subtract the gravitational component of the bispectrum, which is typically $1-2$ orders of magnitude larger than the primordial component (assuming $f_{\rm NL} \sim 1$).}

The final ingredient needed to evaluate the Fisher matrix for $f_{\rm NL}$ is an expression for the dark ages 21cm power spectrum, which appears in Eq.~\ref{eq:deltaB}. This is given by 
\begin{equation}
    P_{\rm HI}(k,\mu,z) = \left(\alpha + \overline{T}_{\rm HI}\,\mu^2\right)^2\,P_{\delta_b}(k) + P_N(k, z),
\end{equation}where $P_{\delta_b}$ is the linear power spectrum of baryon fluctuations.

\subsection{Model of the interferometer noise power spectrum}

The last factor, $P_N$, is the power spectrum of the instrumental noise. \edit{This term describes the noise variance on each Fourier mode} measured by the radio array. This sets the noise level on each triangle configuration that contributes to the bispectrum.

Following the notation of \cite{2020JCAP...11..052K}, the noise power spectrum for an interferometer can be calculated from the radiometer equation as
\begin{equation}
P_N(k, z) = T_{\rm sys}^2 \lambda \frac{(1+z) \chi^2(z)}{H(z)} \left ( \frac{\lambda^2}{A_{\rm eff}} \right)^2 \frac{1}{N_{\rm pol}\, n_b(\vec{u})\, t_{\rm tot}} \frac{S_{\rm area}}{\Omega_{\rm FOV}}, \label{eq:PN}
\end{equation}
where $T_{\rm sys} = T_{\rm sky} + T_{\rm inst} \approx T_{\rm sky}$ is the system temperature, $H(z)$ and $\chi(z)$ are the expansion rate and comoving radial distance to redshift $z$, $\lambda$ is the observed wavelength of the redshifted 21cm (i.e. $\lambda = 0.211 (1+z)\,{\rm m}$), $A_{\rm eff}$ is the effective area of each receiving element of the radio interferometer, $\Omega_{\rm FOV}$ is the solid angle of the beam pattern of each element, $N_{\rm pol} = 1$ or $2$ is the number of available receiver polarisation channels, and $t_{\rm tot}$ and $S_{\rm area}$ are the total observing time for the survey and the total survey area, respectively. We model the sky temperature as $T_{\rm sky} \approx 5000\,{\rm K}\, (\nu / \nu_{\rm ref})^{-2.5}$, where $\nu_{\rm ref} = 50~{\rm MHz}$. \edit{Finally, $n_b(\vec{u})$ is the number density} of interferometer baselines in a region of the $uv$-plane, where $\vec{u}$ is a 2D vector that can be mapped to a transverse Fourier mode on the sky, $\vec{k}_\perp$. 

\begin{figure}[!th]
\centering
\includegraphics[width=0.8\linewidth]{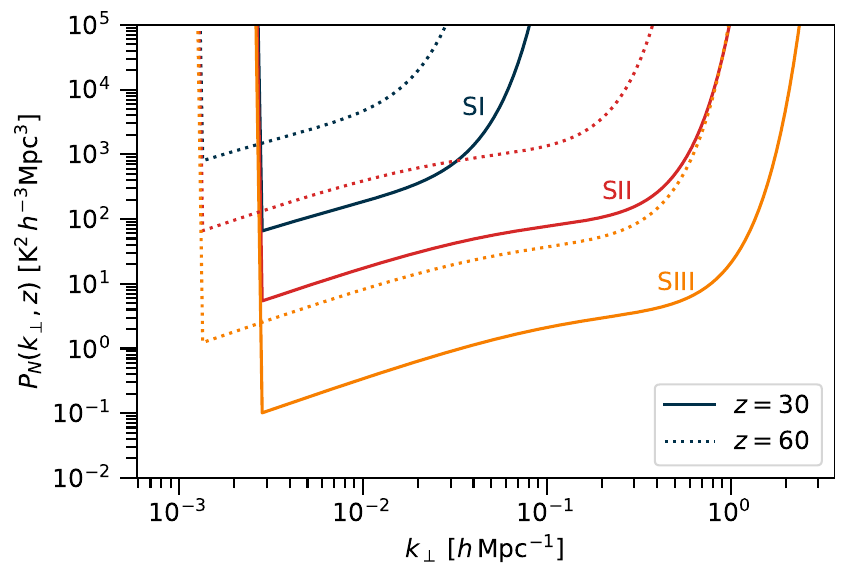}
\caption{Noise power spectrum $P_N(k_\perp,z)$, as given in Eq.~\ref{eq:PN}, assuming $t_\mathrm{tot} = 22,000$~hours of observation time (equivalent to 5 years with a $\sim$50\% duty cycle). The curves correspond to the three types of experiment: Stage~I (blue), Stage~II (red), and Stage~III (orange), with solid lines denoting redshift $z=30$ and dotted lines $z=60$. \edit{The increase in array size gains between one and two orders of magnitude in sensitivity between each stage. For comparison, the model considered by \cite{2023arXiv230508593M} has a dimensionless power spectrum of $\Delta^2 = 0.44$~mK$^2$ at $k=0.1$~Mpc$^{-1}$ and $z=51$ (about $0.025\,{\rm K}^2$ $h^{-3} {\rm Mpc}^3$, i.e. below the detection threshold even for Stage III with these figures).}}
\label{fig:noise}
\end{figure}

\edit{Fig.~\ref{fig:noise} shows the noise power spectra of the three stages of experiment, using the idealised baseline distributions discussed in Sect.~\ref{sec:config}. As a reference survey timescale, we assume $t_{\rm tot} = 22,000$ hours of observing time, which crudely represents an efficient 5-year survey with a 50\% duty cycle, e.g. to account for flagging of data when the Sun is up or bright sources are in the sidelobes. The survey area is assumed to be 20,000~deg$^2$ in each case, with a field of view of $\Omega_{\rm FOV} \sim \pi$ steradians (10,300~deg$^2$). For a dipole-like antenna, the effective area is $A_{\rm eff} = \lambda^2 G / (4 \pi)$, where we assume a slightly enhanced gain of $G \approx 2$. The sharp cutoff at low $k_\perp$ is due to the assumed minimum baseline length of 20m, although lower $k_\perp$ could potentially be recovered if calibrated zero-spacing (autocorrelation) data can be obtained. There is a clear redshift dependence, with a shift to lower $k_\perp$ as $z$ increases due to the frequency dependence of the fringe pattern of each baseline, and a shift to higher noise power as $z$ increases (frequency decreases) due to the increase in system (sky) temperature. As the array increases in size, the number of baselines increases on all scales, increasing the sensitivity by an order of magnitude or more between each generation. Larger arrays also have longer maximum baseline lengths, and so reach a higher effective maximum $k_\perp$.}

\edit{In what follows, we will assume that the noise power spectrum is sub-dominant, i.e. that we are in the sample variance-dominated limit. Reaching this limit would be a major technical feat, involving a very long survey duration, excellent control over systematic effects, calibration errors, and so on. It is nevertheless useful to consider this limit as giving the best constraints that could possibly be achieved with a given array configuration.}

\section{Results}\label{sec:results}

In this section, we present Fisher matrix forecasts for the $f_{\rm NL}$ parameter measured from the HI brightness temperature bispectrum during the Dark Ages, using the formalism described in Sect.~\ref{sec:fisher}. We consider the three stages of experiment described in Table~\ref{tab:expts}, and show the impact of different assumptions about instrumental/scale cuts related to foreground contamination etc.

\begin{figure}[!th]
\centering
\includegraphics[width=0.8\linewidth]{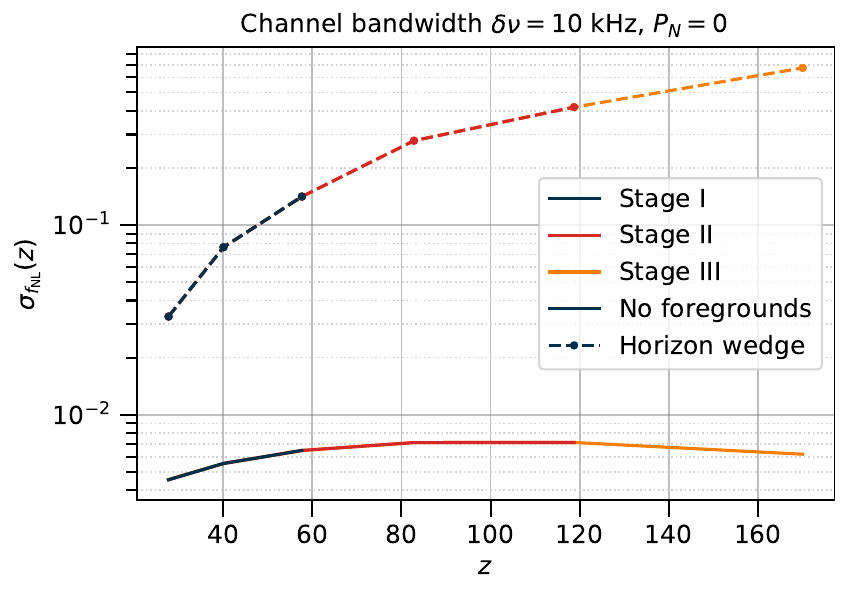}
\caption{Cosmic-variance limited ($P_N=0$) Fisher forecast for the $f_{\rm NL}$ parameter for local-type primordial non-Gaussianities using the three stages of lunar 21cm experiments. The results are shown for the foreground-free case (solid), i.e. $k_\parallel^{\rm min} = 2\pi/V^{1/3}_{z_i}$, and with foregrounds (dashed), i.e. excluding the horizon wedge from the data. The channel width ($\delta \nu$) sets the maximum radial scale $k_{\parallel}^{\rm max}$ included in the analysis. (Note that we use this as the maximum radial scale rather than a non-linear cut-off, as would be more usual for low-$z$ experiments.)}
\label{fig:results}
\end{figure}

For all of the forecasts, we assume a flat $\Lambda$CDM model with parameters $h = 0.674, n_s = 0.965, \sigma_8 = 0.811, \Omega_m = 0.315$, and $\Omega_b = 0.049$ \cite{planck2018}. We restrict ourselves to the redshift range $z = 23-200$ (frequencies between $7 - 60$~MHz) in order to keep the same fitting functions for the HI brightness temperature etc. given in \cite{munoz2015}, and do not marginalise over uncertainties in astrophysical quantities that set the overall scale of the HI signal, or cosmological parameters that set the shape of the non-primordial contributions to the bispectrum. These assumptions will be relaxed in future work, and have been studied in past works, e.g. \cite{munoz2015, 2023arXiv230508593M}. As explained in Sect.~\ref{sec:config}, we also make assumptions about the co-planarity of the baselines, and use a flat-sky limit that also neglects cosmological evolution within discrete redshift bins. In all the cases presented here, the redshift bins are chosen to be sub-bands with $\approx 30\%$ bandwidth at their respective centre frequencies. This is at the upper end of what is plausible if we wish to neglect cosmic evolution within each bin. 

Fig.~\ref{fig:results} shows the forecast 68\% CL error on the $f_{\rm NL}$ parameter for each generation of experiment as a function of redshift. \edit{Under the assumptions we have made -- particularly neglecting the thermal noise contribution -- all three configurations produce similar results in redshift bins where they overlap.}\footnote{In this paper, the following redshift bins are defined such that each comprises around 30\% bandwidth at the respective centre frequencies: (1) $z = 22.7 - 32.8$; (2) $32.8 - 47.3$; (3) $47.3 - 68.0$; (4) $68.0 - 97.6$; (5) $97.6 - 139.9$; (6) $139.9 - 200.2$. Only the first 3 and 5 are present for the Stage I and Stage II experiments respectively.}
\edit{The main difference between the three stages largely comes down to the maximum redshift that each of them can reach. In practice, the Stage I array has about 300 times the noise level of the Stage III array at the same $k_\perp$, and so would need to integrate for a much longer time to reach the same signal to noise ratio.}

\edit{A number of effects contribute to the shape of the curves in Fig.~\ref{fig:results}. The comoving volume of each redshift bin is a key factor in setting the sample variance limit, and increases significantly from low to high redshift. This is tempered by the increasing $T_{\rm sys}$ with redshift. Recall that $f_{\rm NL}$ encodes the amplitude of the primordial bispectrum in the squeezed limit, where there is one low-$k$ leg and two higher-$k$ legs to each triangle. The number of large-scale modes available to form the low-$k$ leg of the squeezed limit triangles also increases with $z$. Redshift-dependent limits on $\sigma(f_{\rm NL})$ of better than $10^{-2}$ can be obtained at $z \gtrsim 20$, surpassing the CMB and low-redshift galaxy surveys. In the case where the foreground wedge region must be excised completely from the data (see Fig.~\ref{fig:scales}), this degrades to around 0.05 at $z \lesssim 50$ and 0.5 at $z \approx 100$, which is still competitive with current constraints. Significantly more evolution in $\sigma(f_{\rm NL})$ with redshift is observed when the wedge is removed, as the size of the wedge region depends on wavelength.}

We stress that these forecasts are optimistic in a number of ways. Most importantly, the secondary/gravitational bispectrum and astrophysical prefactors (i.e. $\overline{T}_{HI}, \alpha, \beta, \gamma$) have been assumed to be perfectly known. In \cite{munoz2015}, these terms were found to be important however, with the astrophysical prefactors exhibiting strong correlations with $f_{\rm NL}$ in their Fisher forecasts. Their bottom-line forecast for $f_{\rm NL}$ was weaker than our prediction, with $\sigma(f_{\rm NL}) \simeq 0.23$ predicted for a large array (other differences, such as their choice of larger 100~kHz channel widths, also contribute to the discrepancy). \edit{We note, however, that relatively mild priors on the astrophysical factors, e.g. from models fitted at lower redshift, should be helpful in breaking the strong correlations, potentially allowing better constraints on $f_{\rm NL}$ to be achieved than presented in \cite{munoz2015}. In this case, our results should be considered to bound the possible values of $\sigma(f_{\rm NL})$ from below, as we are operating in the optimistic sample variance-limited case.}

\section{Conclusions}

In this paper, we reviewed the distinctive properties of the Dark Ages 21cm brightness temperature field as a probe of early universe physics, particularly as encoded by the local-type non-Gaussianity parameter $f_{\rm NL}$. We examined how a lunar radio interferometer could be used to measure the 21cm bispectrum while avoiding severe problems such as ionospheric distortion and radio frequency interference on Earth, and discussed how different instrumental effects contribute scale cuts that limit the number of Fourier modes of the three-dimensional 21cm field that can be recovered. We then went on to define a set of three representative stages of the development of such lunar arrays, beginning with smaller ones with a few hundred antennas spread over a square kilometre, and culminating in a much larger array of a hundred-thousand or more antennas spread over 100~km$^2$. These were used in simple exploratory Fisher matrix forecasts to show how well the 21cm bispectrum, and hence the $f_{\rm NL}$ parameter, could be measured under optimistic assumptions, such as neglecting thermal noise (i.e. obtaining sample variance-limited observations). In essence, our results show the best that a notional lunar 21cm experiment could do in the absence of systematic effects in the parts of the data that remain after a series of scale cuts, and without limits on observing time. 


We found that severe scale cuts that are applied to many ground-based 21cm experiments (at lower redshift) in order to remove foreground contamination did degrade the predicted constraints on $f_{\rm NL}$ substantially, but that under our assumptions the signal could still be measured at a level that is competitive with CMB and galaxy survey experiments.

Overall, we suggest that a staged deployment of lunar 21cm arrays (moving from compact with a few hundred antennas, to widely-distributed with up to a hundred-thousand antennas) should provide a robust path to measuring the local-type non-Gaussianity parameter $f_{\rm NL}$, with the prospect of substantially improved precision compared to ground-based experiments. While there are obvious engineering and cost challenges associated with deploying such arrays on the Moon, we have seen that the cosmological performance of the arrays survive even quite stringent scale cuts to remove foreground contamination and the like. In terms of future work, it would be desirable to go beyond the analytic approach that we have used here, and perform a direct demonstration of 21cm bispectrum recovery on simulated data that include realistic instrumental effects such as foregrounds, calibration errors, antenna variations, and mutual coupling. We have also commented on the need to marginalise over astrophysical uncertainties, which is likely to further reduce the forecasted precision.


\vskip6pt

\ack{\edit{We are grateful to T.~Fl\"oss, D.~Karagiannis, D.~Meerburg, G.~Orlando, and J.~Silk for useful discussions, and two anonymous referees for their helpful comments}. \CG{The code used to produce this work was primarily based on the one provided by \cite{guandalin2022}\footnote{\href{https://github.com/cmguandalin/ClusteringZ-21cm-Bispectrum}{https://github.com/cmguandalin/ClusteringZ-21cm-Bispectrum}}. The coefficients for the 21-cm bispectrum were extracted from \href{https://github.com/JulianBMunoz/Bispectrum21cm}{https://github.com/JulianBMunoz/Bispectrum21cm}}. This work made extensive use of the public code \href{https://github.com/lesgourg/class_public}{\sc{class}} \cite{blas2011}, and the following python packages and libraries: \href{https://numpy.org/}{\sc{numpy}} \cite{harris2020}, \href{https://scipy.org/}{\sc{scipy}} \cite{2020SciPy}, and \href{https://matplotlib.org/}{\sc{matplotlib}} \cite{hunter2007}.}

\funding{We acknowledge the Royal Society for travel support. This result is part of a project that has received funding from the European Research Council (ERC) under the European Union's Horizon 2020 research and innovation programme (Grant agreement No. 948764). C.G. is supported by STFC consolidated grant ST/T000341/1.}


\end{document}